\documentclass[prb,twocolumn,floatfix,showpacs,superscriptaddress]{revtex4}

\usepackage{graphics}

\begin{document}

\title{Oxygen Moment Formation and Canting in Li$_{2}$CuO$_{2}$}
\author{E.M.L Chung} 
\affiliation{Department of Physics, University of Warwick, Coventry, CV4 7AL, UK}

\author{G.J. McIntyre}
\affiliation{Institut Laue Langevin, BP 156, 38042, Grenoble Cedex 9, France}

\author{D. M$^{c}$K. Paul}
\affiliation{Department of Physics, University of Warwick, Coventry, CV4 7AL, UK}

\author{G. Balakrishnan}
\affiliation{Department of Physics, University of Warwick, Coventry, CV4 7AL, UK}

\author{M.R. Lees}
\affiliation{Department of Physics, University of Warwick, Coventry, CV4 7AL, UK}
\date{\today}

\begin{abstract}
The possibilities of oxygen moment formation and canting in the
quasi-1\emph{D} cuprate $\mathrm{Li_{2}CuO_{2}}$ are investigated
using single crystal neutron diffraction at 2 K.  The observed
magnetic intensities could not be explained without the inclusion of a
large ordered oxygen moment of 0.11(1) $\mu_{B}$. 
Least-squares refinement of the magnetic structure
of $\mathrm{Li_{2}CuO_{2}}$ in combination with a spin-density
Patterson analysis shows that the magnetization densities of the Cu
and O atoms are highly aspherical, forming quasi-1\emph{D} ribbons of
localised Cu and O moments. Magnetic structure refinements and
low-field magnetization measurements both suggest that the magnetic
structure of $\mathrm{Li_{2}CuO_{2}}$ at 2 K may be canted. A possible model 
for the canted configuration is proposed.
\vspace{8mm}
\end{abstract}

\pacs{61.12.-q, 75.25.+z, 75.30.-m, 75.40.Cx}

\maketitle

\section{Introduction}
\begin{figure}[t]
\begin{center}
\resizebox{7.5cm}{9cm}{\includegraphics{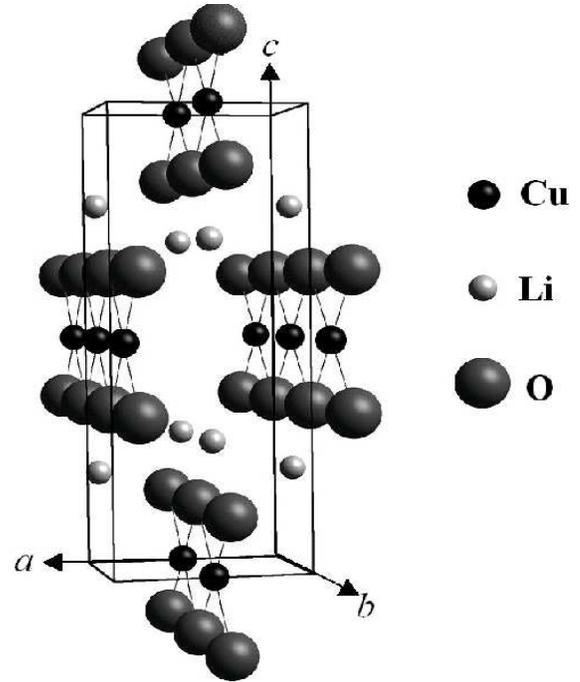}}
\caption{Chemical structure of $\mathrm{Li_{2}CuO_{2}}$ 
showing the CuO$_{2}$ ribbons created along $b$ by an edge-sharing arrangement of
CuO$_{4}$ plaquettes.}
\label{lcouc}
\end{center}
\end{figure}

Quasi-one-dimensional cuprates such as $\mathrm{Li_{2}CuO_{2}}$
exhibit a wide range of 
interesting electronic and magnetic properties. 
As $\mathrm{Li_{2}CuO_{2}}$ possesses 
1D chains of S=$\frac{1}{2}$ copper atoms it serves as a 
useful model system for studies of low dimensional behavior and 
quantum fluctuations in the spin $\frac{1}{2}$ quantum chain. 
Also embedded within the chemical structure of $\mathrm{Li_{2}CuO_{2}}$ 
are sets of parallel Cu-O$_{2}$ ribbons (figure \ref{lcouc}), similar to 
those found in the spin-Peierls compound $\mathrm{CuGeO_{3}}$ and 
chain-ladder system $\mathrm{La_{14-x}Ca_{X}Cu_{24}O_{41}}$. 
$\mathrm{Li_{2}CuO_{2}}$ therefore also 
frequently serves as a model system for investigation of the 
electronic structure and correlations of the Cu-O$_2$ ribbon 
\cite{weht1,neudert1,mizuno1,delley1,degraaf1}. 

The Cu-O networks of $\mathrm{Li_{2}CuO_{2}}$, $\mathrm{CuGeO_{3}}$,
high-$\mathrm{T_{C}}$ materials and many other cuprates can be 
thought of as being constructed from the same basic building block; 
a $\mathrm{CuO_{4}}$ plaquette. In high-$\mathrm{T_{C}}$ compounds, 
the plaquettes are corner sharing forming 2D planes, whilst in 
$\mathrm{Li_{2}CuO_{2}}$ and $\mathrm{CuGeO_{3}}$ they are edge-sharing, 
forming Cu-O ribbons. The magnetic properties of cuprates are 
known to be crucially dependent upon the Cu-O-Cu bond angle linking 
nearest neighboring Cu atoms. When this angle approaches 90$^{o}$ 
the strength of the superexchange 
interaction diminishes and a change in magnetic ordering 
from antiferromagnetic (AFM) to ferromagnetic (FM) occurs 
\cite{goodenough1,kanamori1,anderson1}.
In the context of other cuprates 
$\mathrm{Li_{2}CuO_{2}}$ is unusually placed since it possesses a 
Cu-O-Cu exchange pathway of 94$^{o}$. This angle is close to the 
crossover region between AFM and FM ordering and frustration effects 
resulting from competing interactions are anticipated.

Local spin-density approximation (LSDA) \cite{weht1,delley1} and
quantum chemical \cite{degraaf1} calculations have recently suggested 
that the oxygen
ions may play a more novel and fundamental role in the magnetism of
$\mathrm{Li_{2}CuO_{2}}$ than previously thought. The LSDA
calculations suggest an ordered magnetic moment of 0.92 $\mu_{B}$ per
formula unit, 40\% of which is predicted to be associated with the
`non-magnetic' oxygen sites \cite{weht1}.  Although covalency effects
(e.g. in K$_{2}$IrCl$_{6}$\cite{lynn1}) are not unknown, a moment as
large as 0.2 $\mu_{B}$ associated with each of the oxygen sites would
be unprecedented. A more conservative estimate of the oxygen moment is
provided by quantum chemical techniques, however, these values vary
dramatically (from 0.03 to 0.17 $\mu_{B}$) depending upon which
approximations have been employed to treat the electronic
correlations.

Previous studies on polycrystalline $\mathrm{Li_{2}CuO_{2}}$ include
the determination of the crystal structure via X-rays \cite{hoppe1},
and the observation of an antiferromagnetic ordering temperature
of $\mathrm{T_{N}}\simeq$ 9 K
\cite{ohta1,shreedhar1,okuda1}.  Early neutron diffraction
investigations performed on powder samples at 1.5 K suggested a
collinear arrangement of moments coupled ferromagnetically within the
$a$-$b$ plane, and antiferromagnetically in the $c$-direction
\cite{hoffmann1,sapina1}.  A magnetic moment of 0.96(4) $\mu_{B}$ was
estimated and attributed solely to the magnetic $\mathrm{Cu^{2+}}$
ions \cite{sapina1}. Electron paramagnetic resonance and
antiferromagnetic resonance experiments showed that the easy axis for
the spins lies along the $a$-direction \cite{ohta1}.

More recently, single-crystal muon spectroscopy and 
powder susceptibility measurements have indicated that in weak fields 
the magnetic structure of $\mathrm{Li_{2}CuO_{2}}$ below 2.6 K may be 
canted away from the $a$-axis\cite{staub1,ortega1}. The temperature dependence of the internal fields probed by $\mu$SR also suggests that there may be magnetic moments on both the Cu and the O ions\cite{staub1}.

An excellent test of the reported magnetic structure, and of the validity
of $\mathrm{Li_{2}CuO_{2}}$ LSDA predictions, can be found through
 measurement of the magnetic moment associated with each atom of
$\mathrm{Li_{2}CuO_{2}}$ using single-crystal neutron diffraction. 
By this method it is possible to survey efficiently 
several hundred magnetic and nuclear reflections over a large 
volume of reciprocal space, enabling the magnetic structure of 
$\mathrm{Li_{2}CuO_{2}}$ to be determined with good statistical accuracy.  

In this paper, the growth and characterization of $\mathrm{Li_{2}CuO_{2}}$ 
single crystals, and experimental details of our neutron-scattering measurements 
are first described. In section III the $\mathrm{Li_{2}CuO_{2}}$ magnetic form 
factor and spin-density Patterson functions are introduced.    
Patterson analysis of our observed $\mathrm{Li_{2}CuO_{2}}$ 
magnetic diffraction pattern reveals the existence 
of a large moment associated with the 
oxygen sites. Section IV contains the results of least-squares refinement of 
the magnetic structure of $\mathrm{Li_{2}CuO_{2}}$ with respect to various 
magnetic models. Despite complications 
introduced by a possible slight 
canting of the Cu and O moments, all models consistently 
estimate that the magnitude of the oxygen moment lies
between 0.10 $\mu_{B}$ and 0.12 $\mu_{B}$. 
Lastly, our magnetic model of $\mathrm{Li_{2}CuO_{2}}$ is refined further to 
include the possibility of canting. Section V contains an additional 
investigation of the canted phase, performed using heat capacity and 
single-crystal magnetization measurements. 

\section{Experimental Details}
Single crystals of $\mathrm{Li_{2}CuO_{2}}$ were grown from phase-pure 
polycrystalline $\mathrm{Li_{2}CuO_{2}}$\cite{patat1}
using a four-ellipsoid floating-zone furnace at a rate of 3-6 mm/h 
in 1.3 atm of flowing oxygen. 
The resulting crystalline boules were
faceted and possessed a pronounced 
tendency to cleave along parallel (101) planes.

Room-temperature powder X-ray
diffraction patterns obtained for finely-crushed portions of our 
$\mathrm{Li_{2}CuO_{2}}$ single-crystals showed that the samples were 
of high purity. All peaks were indexed in the orthorhombic 
space group I\emph{mmm} and
our single crystals exhibited all of the expected bulk magnetic 
properties (susceptibility and heat capacity) associated with  
the onset of three-dimensional antiferromagnetic order at 
$T_{N}=9 \pm{0.1}$ K (figure \ref{graphs1}, discussed in section V). 

Neutron diffraction measurements were performed at the high-flux reactor 
at the Institut Laue Langevin (ILL), Grenoble, using the D10 
diffractometer in its four-circle configuration with an offset
Eulerian cradle, 
He-flow cryostat and position-sensitive $80 \times 80$ mm$^{2}$ area 
detector. The 40 mg single crystal platelet of 
$\mathrm{Li_{2}CuO_{2}}$ had (101) flat faces of area $3\times4$  
mm$^{2}$ spaced 0.3 mm apart. 
The positions and intensities of all accessible nuclear and
 magnetic reflections with $k \geq 0$ 
were scanned at 2 K using a constant neutron wavelength of 1.257(5) \AA 
\ to give up to four symmetry equivalents for each unique reflection.  

Background subtraction and peak integration were performed using the
``\emph{minimum in $\sigma(I)/I$ method}''\cite{wilkinson1} where
statistically optimum results are obtained in the integration of
weaker magnetic reflections by using three-dimensional profiles
estimated from stronger neighboring reflections.  The data were
corrected for absorption\cite{coppens1} using an calculated absorption
coefficient of 0.2114 mm$^{-1}$. The transmission factors ranged from
41.7\% to 87.7\% depending on orientation.  Examination of the
intensities of the stronger nuclear reflections suggested that
measurements of the nuclear peaks had suffered significantly from
extinction. The extinction could be accounted for reasonably well in
the structure refinements using a Becker and Coppens type I Lorentzian
model\cite{becker1}, with a crystal mosaicity of 0.03(2)$^{o}$.

\section{The magnetic form factor}

\begin{figure}[t]
\begin{center}
\rotatebox{270}{\resizebox{8cm}{8cm}{\includegraphics{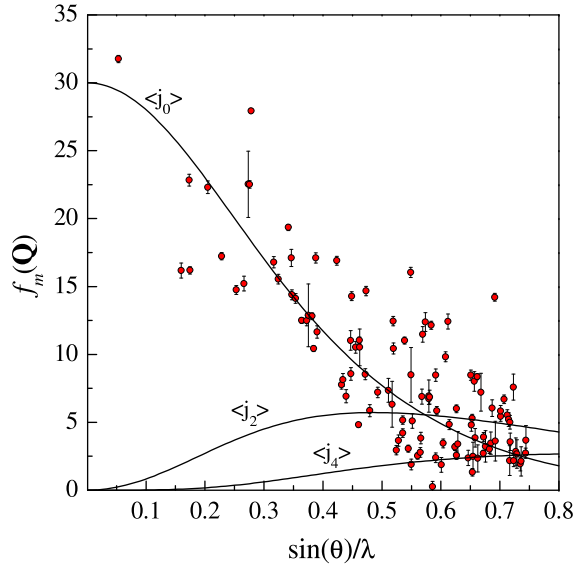}}}
\caption{The magnetic form-factor of $\mathrm{Li_{2}CuO_{2}}$ (markers), calculated from the observed data by assuming a model of $a$-axis Cu moments. The lines 
depict the $<j_{0}>$, $<j_{2}>$ and  $<j_{4}>$ radial integrals
for the free Cu$^{2+}$ ion \cite{itc2} (scaled).}
\label{ff}
\end{center}
\end{figure}

Below $\mathrm{T_{N}}$, magnetic peaks develop at $h$+$k$+$l$= 2$n$+1
(indexed on the basis of the nuclear unit-cell), due to the
antiferromagnetic ordering of Cu moments positioned at (0,0,0) and
($\frac{1}{2}$,$\frac{1}{2}$,$\frac{1}{2}$).  Assuming that all of the
magnetic scattering arises from the Cu atoms the magnetic intensity
can be described by,
\begin{equation}
I_{M}(\textbf{Q}) \propto T^{2} |G(\textbf{Q})f_{m}(\textbf{Q})(1 - e^{-2 \pi i (\frac{h}{2}+\frac{k}{2}+\frac{l}{2})})|^{2}
\label{I1}
\end{equation}
`$T$' is a temperature factor, whose variation with $Q$ at 2 K can be ignored, 
and $f_{m}(\textbf{Q})$ is the atomic magnetic form factor of the Cu site. 
Assuming that the moments are oriented along the $a$-axis the geometrical 
orientation factor, $G(\textbf{Q})$, for this orthorhombic cell is given by,
\begin{equation}
G(\mathbf{Q}) = \frac{\sqrt{(\frac{k}{b})^{2}+(\frac{l}{c})^{2}}}{\sqrt{(\frac{h}{a})^{2}+(\frac{k}{b})^{2}+(\frac{l}{c})^{2}}}
\label{Gfact}
\end{equation}
thus, apart from some constant scale factors the magnetic form factor can be extracted using,
\begin{equation}
f_{m}(\textbf{Q}) \propto \sqrt{I_{M}(\textbf{Q})/|G(\textbf{Q})|^{2}}
\label{ffact}
\end{equation}

The magnetic form factor of $\mathrm{Li_{2}CuO_{2}}$, calculated from the 
measured intensities 
by assuming that magnetic scattering arises exclusively from 
$a$-axis Cu moments is plotted in 
figure \ref{ff}. It can be seen that our measurements deviate radically 
from the $<j_{0}>$ radial integral for the free Cu$^{2+}$ ion \cite{itc2}. 
This is to 
be expected because of the effects of crystalline anisotropy 
on the magnetization density close to the Cu sites.  
A further departure from spherical symmetry may also arise due to the 
presence of a magnetic moment elsewhere in the unit cell, such as on 
the oxygen sites. 

It should be noted that if a large oxygen moment exists, 
or the spins in 
$\mathrm{Li_{2}CuO_{2}}$ are 
significantly canted, the expression for $G(\textbf{Q})$ (provided in 
equation 2) will no longer represent a good approximation for the 
geometrical orientation factor. In this situation, the values estimated for 
$f_{m}(\textbf{Q})$ (using equation 3) could differ considerably from 
those plotted.

\subsection{The Spin-density Patterson Function} 

In principle, the three-dimensional real-space magnetization density 
of the crystal may be obtained directly through Fourier transformation 
(F.T.) of 
the complete magnetic structure factor, \textbf{F}$_{M}({\mathbf{Q}})$. 
The full magnetic structure factor is a vector quantity that carries 
information about the magnitude \emph{and} the direction of the 
spin-density at every point in the unit cell.

\begin{equation}
\mathbf{\rho}_{m}({\mathbf{r}})= {\mathrm{F.T.}}[\mathbf{F}_{M}(\mathbf{Q})]
\label{p1}
\end{equation}

Unfortunately, analysis 
of our neutron diffraction data by this method 
is hindered by the lack of phase information. Since unambiguous knowledge 
of the phase of each of the reflections was unavailable a magnetic 
Patterson approach was adopted\cite{wilkinson2,wilkinson3,wilkinson4}. 
Unlike maximum entropy methods, Patterson analysis requires no phase 
information, or prior knowledge of the positions or the directions of 
the moments. The spin-density Patterson approach treats the magnetic 
structure as being completely unknown.     

The Patterson function was first introduced in 1934, 
and has become an invaluable tool for the determination of 
unknown crystal structures principally via X-rays. 
The non-magnetic Patterson function represents the autocorrelation 
function of the electron density in the unit cell, and is defined 
as the Fourier transform of the squared nuclear structure factor,   
\begin{equation}
P_{N}({\mathbf{u}})= {\mathrm{F.T.}}[|F_{N}(\mathbf{Q})|^{2}]
\label{p2}
\end{equation}
\begin{equation}
P_{N}({\mathbf{u}})= \rho_{n}({\mathbf{r}}) \otimes \rho_{n}({-\mathbf{r}})
\label{p3}
\end{equation}

The nuclear Patterson function, $P_{N}(\mathbf{u})$, provides a map of the 
vectors between \emph{pairs} of atoms in the unit cell. (i.e. in a unit cell 
containing \emph{n} atoms, each atom will form a pair with all atoms, 
including itself, producing $n^{2}$ Patterson peaks).

The relationship between the magnetic Patterson function 
and the unpaired spin-density is not as trivial as the 
situation described for the nuclear Patterson function. 
The dipolar nature of the electron-neutron interaction means 
that the intensity of magnetic scattering is governed by the 
relative orientation of the spins and the scattering vector 
(\textbf{Q}). In terms of \textbf{F}$_{M}$(\textbf{Q}) neutrons 
are scattered by the sample with an intensity given by\cite{halpern1},
\begin{equation}
I_{M}({\mathbf{Q}}) \propto |{\mathbf{F}}_{M}({\mathbf{Q}})|^{2}-|{\mathbf{\hat{Q}}}.{\mathbf{F}}_{M}(\mathbf{Q})|^{2}
\label{haleqn}
\end{equation}
 
Where $\mathbf{\hat{Q}}$ represents a unit vector in the direction of
\textbf{Q}.  Calculation of the spin-density Patterson map by Fourier
transformation of the measured magnetic intensities (equation
\ref{magpateqn}) generates a plot of the pair-correlation function
between magnetic scatterers, equation \ref{reqn},
\begin{equation}
P_{M}(\mathbf{u}) \propto {\mathrm{F.T.}}[|{\mathbf{F}}_{M}({\mathbf{Q}})|^{2}] - {\mathrm{F.T.}}[|{\mathbf{\hat{Q}}}.{\mathbf{F}}_{M}(\mathbf{Q})|^{2}]
\label{magpateqn}
\end{equation}
\begin{eqnarray}
\label{reqn}
P_{M}({\mathbf{u}}) & = & \sum_{i=1}^{3} \sum_{j=1}^{3}(\rho_{i}({\mathbf{r}}) \otimes \rho_{j}(-{\mathbf{r}})) \\
& & - \sum_{i=1}^{3} \sum_{j=1}^{3} {\mathrm{F.T.}}[\hat{Q}_{i}\hat{Q}_{j}] \otimes (\rho_{i}({\mathbf{r}) \otimes \rho_{j}(-{\mathbf{r}}))} \nonumber
\end{eqnarray}

The three spatial components of spin density of the $i$'th and $j$'th atoms 
have been written as summations over $\rho_{i}$ and $\rho_{j}$. $\hat{Q}_{i}\hat{Q}_{j}$ 
describes the product of the magnitudes of the $i$'th and $j$'th components of 
the unit vector $\mathbf{\hat{Q}}$. The symbol $\otimes$ denotes a convolution. 

The second term in equation \ref{reqn} probes the directions of the moments, 
causing the Patterson peaks to become elongated along the direction parallel 
to the vector bisector between spins. Full details of this theory, and references 
relating the uses of the Patterson function in magnetism can be found in 
the references\cite{wilkinson2,wilkinson3,wilkinson4}.

\begin{figure}[t]
\begin{center}
\resizebox{7.2cm}{8.6cm}{\includegraphics{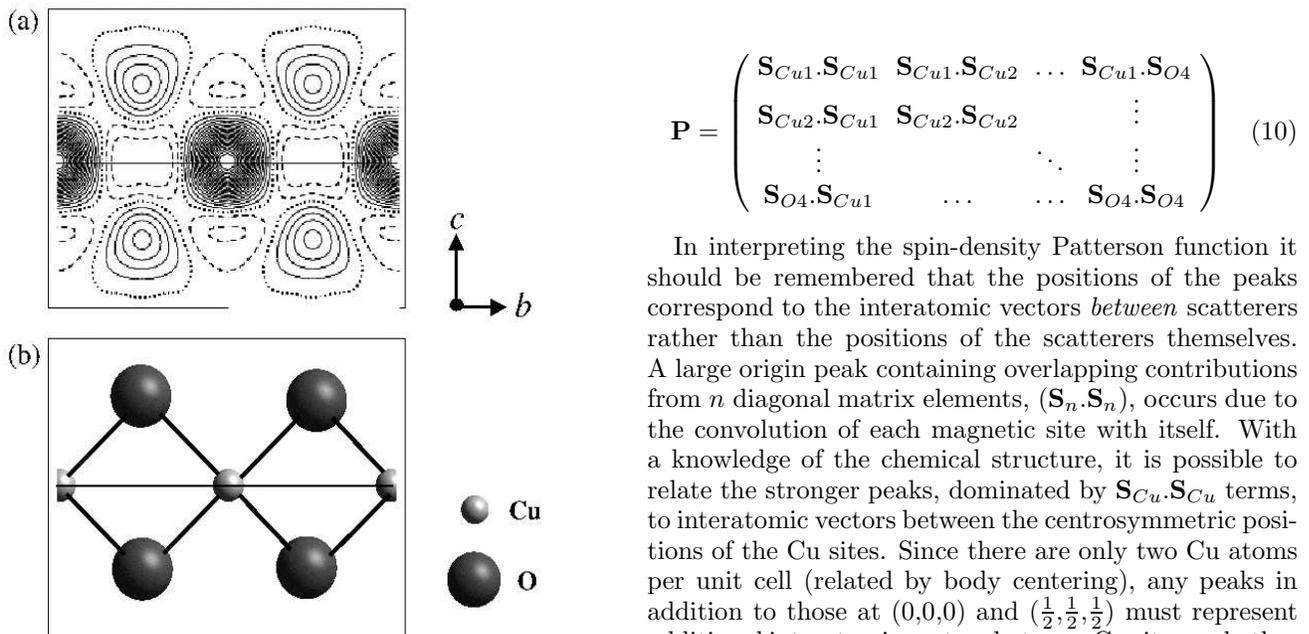}}
\caption{(a) The magnetic Patterson section generated from the observed data 
($|F_{M}(hkl)|^{2}$), viewed as a ($\frac{1}{2}$ $b$ $c$) slice through the 
Li$_{2}$CuO$_{2}$ unit cell. Note that significant magnetic intensity is 
associated with inter-atomic vectors between copper and oxygen sites. 
This confirms the presence of unpaired spins on the oxygen sites, 
see equation \ref{mateqn}. Filled, dotted and dashed lines represent positive, zero and negative 
contours respectively. Contours extend from -5 to 50 in 2.5 steps. 
(b) The corresponding positions of the atoms in the Li$_{2}$CuO$_{2}$ unit cell. }
\label{pat1}
\end{center}
\end{figure}

\subsection{Li$_{2}$CuO$_{2}$ Spin-density Patterson Analysis} 

For the purposes of Patterson analysis, the $\mathrm{Li_{2}CuO_{2}}$ magnetic data 
set of 409 observations was reduced to  a complete set of 111 averaged unique reflections. 
Including the possibility of magnetic oxygen sites there are $n$=6 magnetic atoms per 
chemical unit cell producing $n^{2}$ = 36 Patterson peaks. The heights of the 36 peaks 
may be represented by a symmetric matrix $\mathbf{P}$. Each matrix element corresponds 
to the scalar product of pairs of spins at sites $i$ and $j$.  

\begin{equation}
\mathbf{P}=\left(
\begin{array}{cccc}
\mathbf{S}_{Cu1}.\mathbf{S}_{Cu1} & \mathbf{S}_{Cu1}.\mathbf{S}_{Cu2} & \ldots & \mathbf{S}_{Cu1}.\mathbf{S}_{O4} \\
\mathbf{S}_{Cu2}.\mathbf{S}_{Cu1} & \mathbf{S}_{Cu2}.\mathbf{S}_{Cu2} & & \vdots \\
\vdots & & \ddots & \vdots \\
\mathbf{S}_{O4}.\mathbf{S}_{Cu1} & \ldots & \ldots & \mathbf{S}_{O4}.\mathbf{S}_{O4}
\end{array}
\label{mateqn}
\right)
\end{equation}

In interpreting the spin-density Patterson function it should be remembered that the 
positions of the peaks correspond to the interatomic vectors \emph{between} scatterers 
rather than the positions of the scatterers themselves. A large origin peak containing 
overlapping contributions from $n$ diagonal matrix elements, ($\mathbf{S}_{n}.\mathbf{S}_{n}$), 
occurs due to the convolution of each magnetic site with itself. With a knowledge of 
the chemical structure, it is possible to relate the stronger peaks, dominated by 
$\mathbf{S}_{Cu}.\mathbf{S}_{Cu}$ terms, to interatomic vectors between the centrosymmetric 
positions of the Cu sites. Since there are only two Cu atoms per unit cell 
(related by body centering), any peaks in addition to those at (0,0,0) and 
($\frac{1}{2}$,$\frac{1}{2}$,$\frac{1}{2}$) must represent additional interatomic vectors 
between Cu sites and other magnetic scatterers.

\begin{figure}[t]
\begin{center}
\resizebox{6.5cm}{10.5cm}{\includegraphics{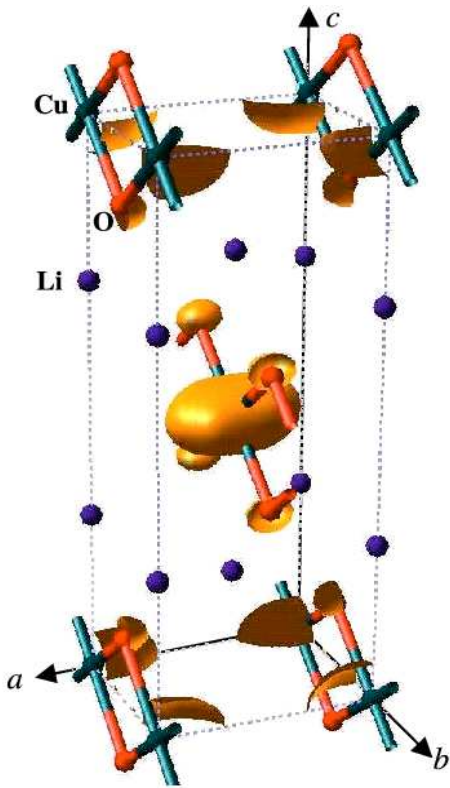}}
\caption{Three-dimensional Patterson map superimposed upon the $\mathrm{Li_{2}CuO_{2}}$ unit cell. 
Contours at (0,0,0) and ($\frac{1}{2}$,$\frac{1}{2}$,$\frac{1}{2}$) 
are fixed at values of  +45 and -45 respectively, (illustrating the reversal in sign between 
spins at (0,0,0) and ($\frac{1}{2}$,$\frac{1}{2}$,$\frac{1}{2}$)).}
\label{pat2}
\end{center}
\end{figure}

In the observed Patterson map (figures \ref{pat1} and \ref{pat2}) additional peaks 
are observed at positions that do not correspond to inter-atomic Cu vectors. 
Comparison with the positions of the atoms in the unit cell shows that a considerable 
number of the magnetic scatterers are closely associated with the crystallographic 
positions of the oxygen atoms. (These peaks are dominated by $\mathbf{S}_{Cu}.\mathbf{S}_{O}$ terms).  
Asymmetry in the magnetic Patterson peaks in the immediate vicinity of the O sites 
(as viewed along the $a$-axis) is in good agreement with the exchange 
potential ($V_{\uparrow}-V_{\downarrow}$) predicted for the 
ferromagnetic Cu-O chain by LSDA calculations\cite{weht1}. 

A three-dimensional Patterson plot of the observed data is 
provided in figure \ref{pat2}. A reversal in sign between CuO$_{4}$ 
units positioned at (0,0,0) and ($\frac{1}{2}$,$\frac{1}{2}$,$\frac{1}{2}$) 
is observed. Elongation of the peaks along the $a$-direction is consistent 
with the direction of the bisector of the spins being parallel to the 
$a$-axis, which does not exclude the possibility of canting.

\section{Least-squares Refinement}

Although Patterson analysis has proved useful for locating the distribution of spin-density 
in the $\mathrm{Li_{2}CuO_{2}}$ unit cell, an estimate has not so far been obtained for the 
magnitudes and directions of the spins. 
A quantitative estimation of the magnitude of the moments in 
$\mathrm{Li_{2}CuO_{2}}$ can be 
gained through least-squares refinement of the magnetic data set with respect to an 
appropriate magnetic model (or models).

Before refining the magnetic structure of $\mathrm{Li_{2}CuO_{2}}$ the nuclear structure was 
verified using the CCSL program SFLSQ \cite{ccsl}. The nuclear data set consisted of 445 
reflections ($h$+$k$+$l$=2$n$) measured at 2 K.
The atomic positions and thermal parameters resulting from 
refinement of the data are provided in Table \ref{tabnuc}. Unit cell parameters at 2 K 
were refined to be $a$=3.661(4), $b$=2.866(7), and $c$=9.397(8) \AA. The high value 
of $\chi^{2}$=17.9 is likely to be due to extinction effects that have not been perfectly 
corrected by the Becker-Coppens procedure.

\begin{table}[h]
\caption{$\mathrm{Li_{2}CuO_{2}}$ atomic positions and thermal parameters 
measured at 2 K, obtained through the refinement of 9 basic variables, 
$\chi^{2}$=17.9, $R(F)$=2.15. ($R(F)=\sum{|F_{0}-F_{C}|}/\sum{|F_{0}|}$ 
and $\chi^{2}=\frac{1}{n-p} \sum{w(F_{O}^{2}-F_{C}^{2})}$, where $n$ is the number 
of points, 
$p$ the number of parameters and $w$ the weighting.) Thermal parameters, 
(assumed to be isotropic for Li and Cu, and anisotropic for O,) correspond 
to the expression $\exp[-({h^{2}\beta_{11}+k^{2}\beta_{22}+l^{2}\beta_{33}})]$.}
\begin{ruledtabular}
\begin{tabular}{lllll}
Atom & X  & Y & Z & Thermal parameters\\ 
\hline
Li & 0 & 0 & 0.7132(2) & $\beta_{11}=\beta_{22}=\beta_{33}=0.326(20)$\\
\hline
Cu & 0 & 0 & 0         & $\beta_{11}=\beta_{22}=\beta_{33}=0.095(10)$\\
\hline
  &  & & & $\beta_{11}=0.279(14)$\\
O   & 0.5    & 0  & 0.6421(5)          & $\beta_{22}=0.161(16)$\\
   &     &   &           & $\beta_{33}=0.229(15)$\\
\end{tabular} 
\end{ruledtabular}
\label{tabnuc}
\end{table}

Refinement of the magnetic structure of $\mathrm{Li_{2}CuO_{2}}$ was performed using the 
CCSL programs MAGLSQ and MMPLSQ. The atomic co-ordinates, 
thermal displacement parameters and overall scale factor 
were fixed to the values found from refinement against the 
nuclear reflections (table \ref{tabnuc}). Calculated
magnetic form factors were assumed for the copper\cite{itc2} and 
oxygen\cite{itc2,Oformfactor} sites. As preliminary 
refinements showed that extinction corrections for the magnetic peaks 
were negligible, the magnetic reflections could be grouped and averaged.

\begin{figure}[t]
\begin{center}
\resizebox{6.3cm}{13cm}{\includegraphics{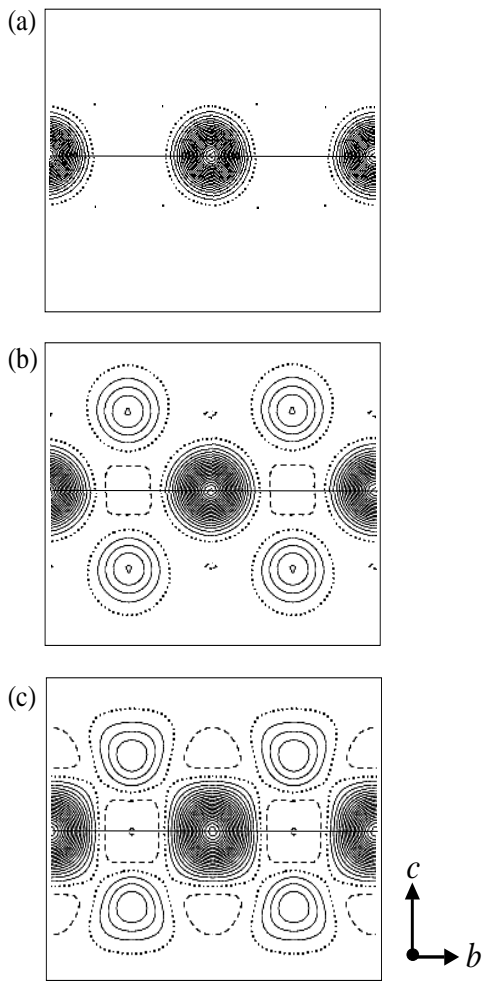}}
\caption{Magnetic Patterson sections taken through the Cu-O ribbon, generated 
from least-squares refined model data. Figure \ref{pat3}(a) spherical $a$-axis Cu moments. 
Fig. \ref{pat3}(b) spherical $a$-axis Cu and O moments. Fig. \ref{pat3}
(c) non-spherical $a$-axis Cu and O moments. For comparison, the observed 
Patterson map is provided in fig.\ref{pat1}.}
\label{pat3}
\end{center}
\end{figure}

For completeness, refinement of the magnetic structure of $\mathrm{Li_{2}CuO_{2}}$ 
was first carried out by assuming  a model based upon previous reports - 
an antiferromagnetic arrangement 
of $a$-axis Cu moments \cite{hoffmann1,sapina1,ohta1}. Refinements based on this model 
(with spherical form-factors) estimated an $a$-axis Cu moment of 0.938 $\pm$ 0.024 $\mu_{B}$. 
This value is close to the 0.96 $\mu_{B}$ estimated from powder neutron diffraction\cite{sapina1}. 
However, considering that there are no extinction effects associated with the magnetic 
reflections, the model provides a poor fit to the data ($\chi^{2}$ = 45), see 
figure \ref{pat3}(a). A number of model 
structures containing canted copper moments were also investigated but none were found that 
enhanced the refinement. Tilting of the Cu moments away from the $a$-axis in itself was 
not sufficient to explain the observed magnetic intensities.

A more satisfactory fit to the data ($\chi^{2}$= 31) could be 
achieved through the inclusion of an ordered oxygen moment, 
see figure \ref{pat3}(b). Comparison of the Patterson map generated 
using this model with the observed 
Patterson map in figure \ref{pat1}(a) shows that the magnetization 
densities of the Cu and O sites are asymmetric in the $b$-$c$ plane. 
This strongly indicates the importance of aspherical multipolar contributions. 

Asymmetry in the magnetization densities of the Cu and O sites was modeled 
using linear combinations of spherical harmonics. The angular dependence of the 
magnetization density of the centro-symmetric Cu sites was modeled using an 
approximation of the form of Equation \ref{cuff}, where the radial integrals 
$<j_{0}>$, $<j_{2}>$ and $<j_{4}>$ (as shown in figure 2) are now modified to possess an angular dependence expressed in terms of Bessel functions $Y[l,m]$. The non-centrosymmetric O angular dependence was modeled using an approximation in the form of equation 
\ref{off}. The coefficients $a_{l,m}$ are fitted parameters determined during the refinement. 

\begin{equation}
\begin{array}{rl}
\rho_{Cu}= & a_{0,0}<j_{0}>+a_{2,0}<j_{2}>Y[2,0]\\
 & +<j_{4}>(a_{4,0}Y[4,0]+a_{4,2^{+}}Y[4,2^{+}])\\
\end{array}
\label{cuff}
\end{equation}

\begin{equation}
\begin{array}{rl}
\rho_{O}= & a_{0,0}<j_{0}>+a_{1,0}<j_{1}>Y[1,0]\\
  &+ <j_{2}>a_{2,2}Y[2,2^{+}]\\
\end{array}
\label{off}
\end{equation}

Sequential introduction of aspherical form factors into our model, 
refined with MMPLSQ, was found to 
improve the fit remarkably. The Patterson section along the Cu-O chain, 
obtained using data calculated by assuming a model containing aspherical 
Cu and O form-factors, is depicted in figure \ref{pat3}(c). Spherical 
harmonics that did not contribute significantly were omitted from the 
final refinement.  

A summary of the results of analysis of the magnetic dataset using a 
variety of starting models is 
presented in table \ref{tabmag}. Spherical harmonics introduced to model 
the magnetization densities of the Cu and O sites are indicated by the 
values of $Y$[$l$,$m$]. (For example, $Y$[0,0] indicates that the 
magnetization density was assumed to be perfectly spherical.)
The magnitude of the O moment, refined to be approximately 0.11 $\mu_{B}$ in size, 
remained stable throughout the refinements. These O moments are coupled to Cu 
moments between 0.91 $\mu_{B}$ and 0.96 $\mu_{B}$ in size. 

\begin{table}[h]
\caption{$\mathrm{Li_{2}CuO_{2}}$ magnetic structure refinements performed on a 
unique set of 111 reflections. Spherical harmonics introduced to model the 
magnetization densities of the Cu and O sites are indicated by the values of 
$Y[l,m]$, (where $l$ and $m$ denote electronic quantum numbers). $\theta$ is the 
angle away from $c$*; $\phi$ is the angle to $a$* in the $a$-$b$ plane.} 
\begin{ruledtabular}
\begin{tabular}{llll}
MODEL & Cu ($\mu_{B}$)  & O ($\mu_{B}$) & $\chi^{2}$ \\ 
\hline
$a$-axis Cu & 0.938 & -- & 44.9\\ 
Cu: $Y$[0,0] & $\pm$0.024 & & \\  
\hline
$a$-axis Cu \& O, & 0.911 & 0.105 & 30.7\\
Cu, O: $Y$[0,0]          & $\pm$0.021 & $\pm$0.012 &  \\
\hline
$a$-axis Cu \& O & & & \\
Cu: $Y$[0,0], $Y$[2,*] \& $Y$[4,*] & 0.960 & 0.100 & 15.7 \\ 
O: $Y$[0,0]  & $\pm$0.017 & $\pm$0.008& \\
\hline
$a$-axis Cu \& O, & & & \\ 
Cu: $Y$[0,0], $Y$[2,*] \& $Y$[4,*]&0.963 &0.105 & 11.1\\
O: $Y$[0,0], $Y$[1,0] \& $Y$[2,2$^{+}$] &  $\pm$ 0.014 & $\pm$0.007 & \\ 
\hline
Canted($\theta$, $\phi$): & & & \\ 
Cu(76$^{o}$,0$^{o}$), O(116$^{o}$,0$^{o}$), & & & \\ 
Cu: $Y$[0,0], $Y$[2,*] \& $Y$[4,*]& 0.914 & 0.116 & 8.75 \\
O: $Y$[0,0], $Y$[1,0] \& $Y$[2,2$^{+}$]& $\pm$0.006 & $\pm$0.004 & \\ 
\end{tabular} 
\end{ruledtabular}
\label{tabmag}
\end{table}

\subsection{Canting}
Although asphericity can be seen to play an important role in describing 
the magnetism of $\mathrm{Li_{2}CuO_{2}}$, the high value of $\chi^{2}$=11.1 indicates 
that further information can be extracted from the data. Difficulties encountered during 
refinement of the magnetic structure of $\mathrm{Li_{2}CuO_{2}}$  
may be explained by mounting evidence suggesting that below T = 2.6 $\pm$ 
0.2 K the moments are slightly canted\cite{staub1,ortega1}. Other than the transition temperature, 
little is known about the nature of the canted phase, except that the canting angle is 
likely to be in excess of 0.7$^{o}$ \cite{staub1}. 

\begin{figure}[t]
\begin{center}
\resizebox{7cm}{8cm}{\includegraphics{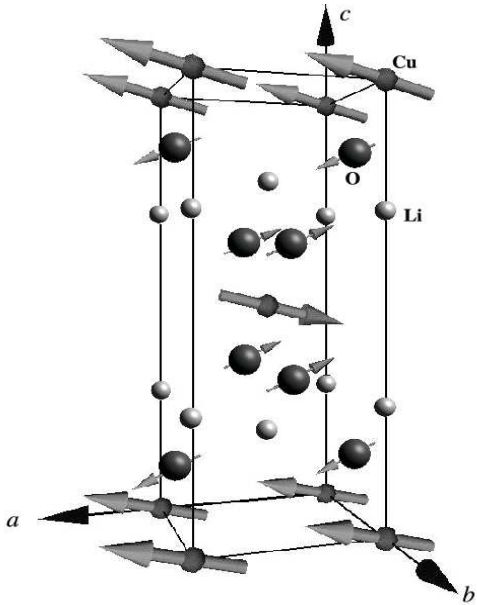}}
\caption{The antiferromagnetically canted structure found to be most consistent with the observed data.}
\label{cant}
\end{center}
\end{figure}

\begin{figure}
\begin{center}
\resizebox{8 cm}{18.5cm}{\includegraphics{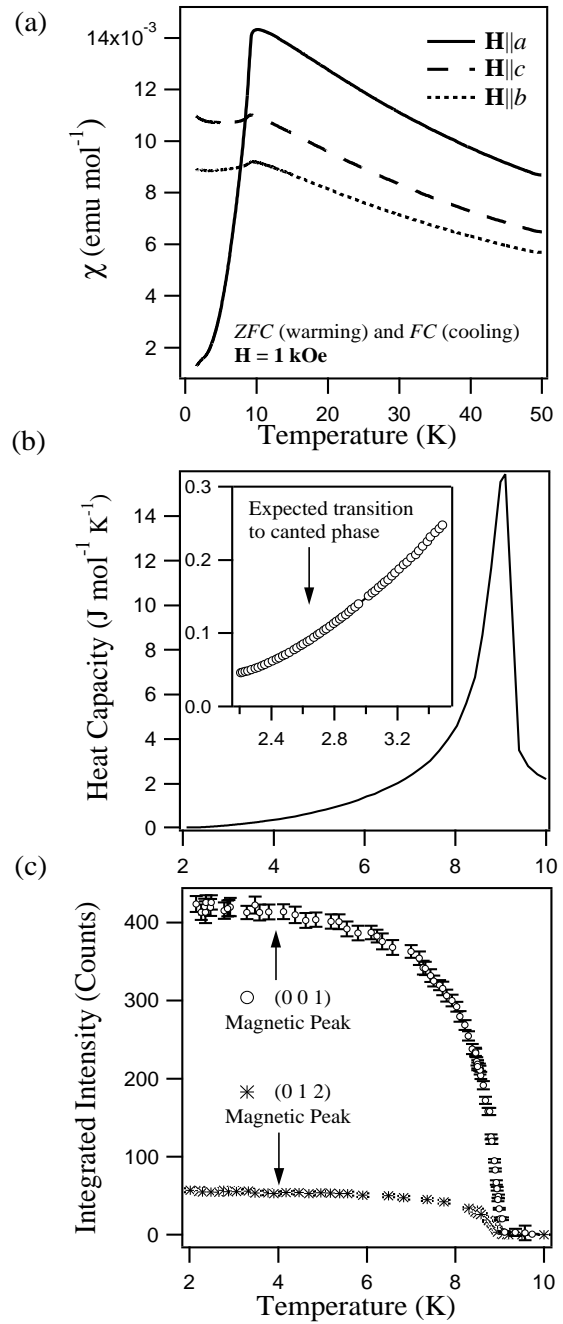}}
\caption{(a) $\mathrm{Li_{2}CuO_{2}}$ single-crystal susceptibility
curves measured in \textbf{H}=1 kOe with \textbf{H} applied parallel
to the $a$, $b$ and $c$-axes. The crystal was warmed to room
temperature and then zero-field-cooled (\emph{ZFC}) between
measurements. Each curve contains data obtained during warming
(\emph{ZFC} data) then cooling of the sample in field (\emph{FC}
data). In 1 kOe no low temperature magnetic transition is
observed. (b) $\mathrm{Li_{2}CuO_{2}}$ zero-field heat capacity
data. Data in the temperature range of the possible transition to the
canted phase are shown in the inset. No anomaly in the heat capacity
in the region of the transition is observed. (c) Integrated intensities
of the (001) and (012) magnetic Bragg peaks measured on warming.}
\label{graphs1}
\end{center}
\end{figure}

In the present study we consider the possibility of an
\emph{antiferromagnetic} canted arrangement of moments. (In an
antiferromagnetically canted arrangement the moments are tilted so
that the net magnetization in all directions remains zero, and no
magnetic intensity is contributed to the positions of the nuclear
peaks.) This restriction seems reasonable considering the absence of a
second transition in the temperature dependence of the (012) and (001) peaks,
figure \ref{graphs1}(c). To the best of our knowledge no evidence
exists in support of a ferromagnetically canted arrangement of moments
in zero-field. Canting of the moments towards the $c$-axis, with the
oxygen moments tilted in the opposite direction to those of the
copper, was found to provide the best overall improvement to our
refinements.

A diagram of our proposed model is provided in figure
\ref{cant}. Adjacent Cu and O moments are coupled ferromagnetically in
the $a$-direction with a small antiferromagnetically coupled $c$-axis
component. The magnitude and direction of the moments was varied until
the lowest value of $\chi^{2}$ was obtained. ($\chi^{2}$=8.75 with 111
averaged reflections and $\chi^{2}$=6.65 with the full set of 232
reflections.)  The resulting model consisted of 0.91 $\mu_{B}$ Cu
moments canted in the $a$-$c$ plane at an angle of $\theta$=76$^{o}$
to the $c$-axis. Oxygen moments approximately 0.12 $\mu_{B}$ in size
are counter-canted in the $a$-$c$ plane at an angle of
$\theta$=116$^{o}$. Although this result suggests that the magnetic structure of $\mathrm{Li_{2}CuO_{2}}$ is canted, it should be remembered that a small reduction in $\chi^{2}$ might also be expected as a side-effect of increasing the number of fitting parameters. 

The ordered $a$-axis moment per formula unit was 1.06 $\mu_{B}$. This
is consistent with the 0.96(4) $\mu_{B}$ estimated from powder neutron
diffraction \cite{sapina1}, and the spin-only value of the free Cu ion
(1 $\mu_{B}$). (For most \emph{d}-electron transition metal oxides the orbital momentum is expected to be quenched.)

There are also several published estimates for the effective moment per Cu atom obtained from susceptibility measurements. Large discrepancies between these estimates, which vary between 1 $\mu_{B}$ and 2.3 $\mu_{B}$\cite{okuda1,ortega1,boehm1}, are probably due to the combined effects of a preferred [101] orientation of polycrystalline grains, and inappropriate fitting of the data to a Curie-Weiss law. High quality single crystal $\mathrm{Li_{2}CuO_{2}}$ susceptibility measurements clearly show that the Curie-Weiss regime (T$>>J$) does not commence until temperatures well in excess of 300 K\cite{mizuno1}. 

\section{Magnetisation data}

In order to investigate further the nature of the possible canting in 
$\mathrm{Li_{2}CuO_{2}}$, single-crystal magnetization measurements 
were obtained between 1.6 and 50 K using an Oxford Instruments vibrating 
sample magnetometer (VSM). The moderately moisture sensitive
$\mathrm{Li_{2}CuO_{2}}$ crystals were polished to remove surface
contamination and then oriented using X-ray Laue photography. 
By this method the orientation of the crystal
axes relative to the applied field could be controlled to within
$\mathrm{\pm 5^{o}}$. 
The magnetic properties of three separate crystal specimens weighing 
38 mg, 154 mg, and 86 mg were investigated. The latter crystal was 
deliberately exposed to moisture in order 
to identify any magnetic impurities arising from surface decomposition; none were observed.

Susceptibility curves obtained along the $a$, $b$ and $c$-axes respectively, 
in an applied field of $\mathbf{H}$=1 kOe, are presented in figure \ref{graphs1}(a).  
These results are in excellent agreement with previous reports 
\cite{mizuno1,boehm1}. A sharp drop in the susceptibility associated with
easy $a$-axis antiferromagnetic ordering is observed at T$_{N}$ = 9 $\pm$ 0.1 K. 
Zero-field heat capacity data is presented in figure \ref{graphs1}(b). Special 
care was taken to follow the heat capacity in the vicinity of the expected low 
temperature transition,  \ref{graphs1}(b) (inset). The integrated intensities 
of the (012) and (001)magnetic Bragg peaks are shown in figure \ref{graphs1}(c). None of the measurements presented 
in figure \ref{graphs1} show any visible anomalies in the low-temperature region between 2 and 4 K. 

\begin{figure}[t]
\begin{center}
\resizebox{7.5cm}{14cm}{\includegraphics{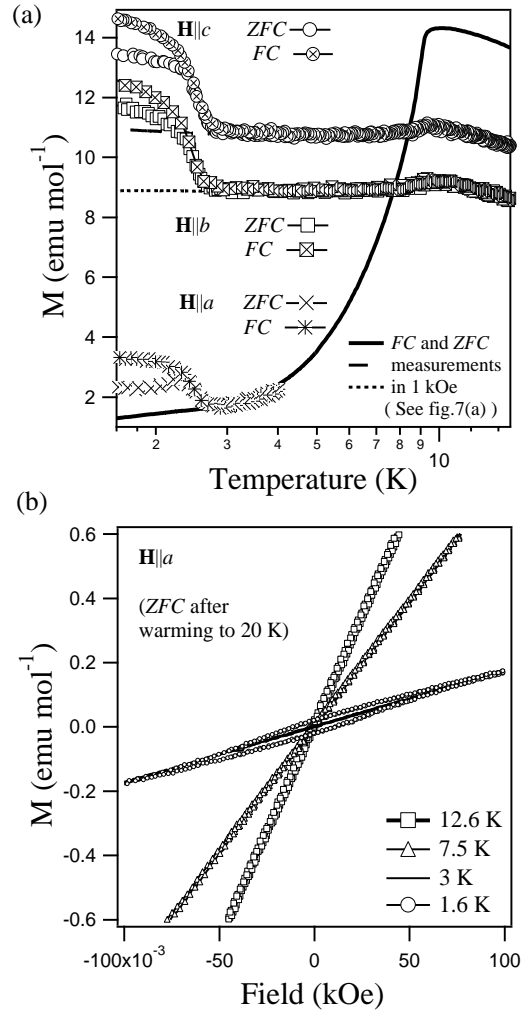}}
\caption{(a) $\mathrm{Li_{2}CuO_{2}}$ susceptibility curves measured in 
$\mathbf{H}$=20 Oe along the $a$, $b$ and $c$-axes. Between each measurement 
the sample was warmed to at least 20 K and then zero-field cooled (\emph{ZFC}). 
Data was then collected during warming in a field of 20 Oe (\emph{ZFC}), and then on cooling in a field of 20 Oe (\emph{FC}). 
The low-field data has been scaled to coincide with the data measured in 
$\mathbf{H}$=1 kOe (shown as lines), see fig. \ref{graphs1}(a). (b) \textbf{M}-\textbf{H} loops 
obtained at 1.6 K, 3 K, 7.5 K and 12.6 K with \textbf{H} applied parallel to the $a$-axis.}
\label{graphs2}
\end{center}
\end{figure}

VSM data obtained between 1.6$ < $T$ < $20 K in a weak field of $\mathbf{H}$ $\simeq$ 20 Oe 
are presented in figure \ref{graphs2}(a). We find that the response of the 
bulk crystal to a small applied magnetic field changes abruptly at 
approximately 2.6 K, producing a small ferromagnetic component in the 
direction of the field. Below 2.6 K a small ferromagnetic hysteresis 
was also observed along each of the 
measured directions, between -100$<\mathbf{H}<$100 Oe, figure \ref{graphs2}(b). 

Although it is possible that the rise in magnetisation could be due to the ordering of a ferromagnetic impurity, the low temperature transition was consistently observed in all three of our samples, and has been reported previously in powder susceptibility\cite{ortega1} and $\mu$SR data\cite{staub1}. Assuming that the observed transition is an intrinsic property of the crystal, we find that at $\sim$ 2.6 K the Cu moments, the O moments, (or both), become free to cant toward the applied magnetic field. The observation of ferromagnetic hysteresis when $\mathbf{H}$ is parallel to the $a$-axis (figure \ref{graphs2}(b)), indicates that the moments may not be perfectly aligned along $a$, even above 2.6 K or in zero field.

\section{Discussion}

To conclude, we have performed detailed measurements of the magnetic
form-factor of a single crystal of $\mathrm{Li_{2}CuO_{2}}$ at 2
K. Patterson analysis of a complete set of unique reflections derived
from the observed magnetic diffraction pattern reveals the presence of
an effective oxygen moment, which appears to be large and
well-localised. Least squares refinement of the magnetic structure in
conjunction with the spin-density Patterson analysis reveals that the
magnetization densities surrounding the Cu and O sites are strongly
aspherical.  The magnitudes of the O and Cu moments remained
approximately stable throughout our refinements. This allows us to
estimate that the magnitude of the O moment lies between 0.10
$\mu_{B}$ and 0.12 $\mu_{B}$. These O moments are coupled to Cu
moments that are approximately 0.92 $\mu_{B}$ - 0.96 $\mu_{B}$ in
size. Despite the magnitude of the oxygen moment being substantially
lower than that estimated by LSDA calculations\cite{weht1}, our
observations are otherwise in good qualitative agreement with LSDA
predictions for the exchange potential of the FM chain. The size of
the oxygen moment computed using the quantum chemical `difference
dedicated configuration interaction' technique for
$\mathrm{Li_{2}CuO_{2}}$ is closer to that measured in this experiment
\cite{degraaf1}.
 
Although large covalency effects have been observed previously in
structures such as $\mathrm{K_{2}IrCl_{6}}$\cite{lynn1}, reports of
moment formation in oxides are extremely rare.  The oxygen moment in
$\mathrm{Li_{2}CuO_{2}}$ is one of the largest so far recorded, rivaled in
size only by the 0.1 $\mu_{B}$ moment suggested to reside on the
oxygen sites in the ferromagnetic manganite
$\mathrm{La_{0.8}Sr_{0.2}MnO_{3}}$\cite{pierre1}. (In the 
polarized neutron study of the manganite an oxygen moment was identified, but attempts
to refine its magnitude from the measured structure factors proved
unsuccessful.)  In other examples of oxygen moment formation the
transfer of moment from the magnetic ion to the oxygen site has been
shown to be much smaller. For example, the only confirmed oxygen
moment in a cuprate system is of 0.02 $\mu_{B}$ in the chain-ladder
compound $\mathrm{La_{14-x}Ca_{x}Cu_{24}O_{41}}$\cite{matsuda1}.

Besides the discovery of the O moment in $\mathrm{Li_{2}CuO_{2}}$, our
observations also seem consistent with the notion of canting. A ferromagnetic-like rise in the magnetization of our single-crystal samples was verified by low-field
($\mathbf{H} =$ 20 Oe) magnetization measurements to occur at T = 2.6
$\pm$ 0.2 K.  However, since many possible configurations of canted Cu and O
moments may be compatible within error with the observed magnetic diffraction
pattern, it is difficult to reach any reliable conclusions about the
nature of the canting. Having restricted our refinements to
antiferromagnetic canted configurations, our most successful model
of the observed data consists of $\sim$ 0.91
$\mu_{B}$ Cu moments canted in the $a$-$c$ plane at an angle of
$\theta$=76$^{o}$ to the $c$-axis. Nearest neighboring ~ 0.12 $\mu_{B}$ O moments are ferromagnetically coupled to the Cu along $a$, but
counter-canted at an angle of $\theta$=116$^{o}$ from $c$, (see also
figure \ref{cant}). 

Previous researchers \cite{weht1} have suggested that the Cu $d_{yz}$
and O $p_\sigma$ orbitals, which extend along the magnetic ribbons,
are energetically isolated and strongly hybridized. Covalency effects
serve to polarize the neighboring oxygen ions, resulting in a large
net transfer of the Cu moment to the oxygen sites. In
$\mathrm{Li_{2}CuO_{2}}$ we find that the Cu moment is large and
well separated from that of the Cu. Counter-canting of adjacent oxygen
moments seems inconsistent with the idea of a purely
polarization-induced oxygen magnetization, and it is possible that the
true magnetic structure of $\mathrm{Li_{2}CuO_{2}}$ may be
frustrated, ferromagnetically canted or more complicated than our proposed
model. (A ferromagnetically canted configuration of moments would have
contributed a slight intensity to the nuclear peaks and was therefore
difficult to refine using our existing data.) In order to determine
the nature configuration of any canted phase with more confidence further
experiments are required.

\section{Acknowledgements}
The authors thank Julie Staunton, Martin Boehm, Jim Hague, Clive
Wilkinson and Jane Brown for useful discussions, and the Engineering and
Physical Sciences Research Council, UK, for financial support.

\bibliography{lco}

\begin{thebibliography}{31}
\expandafter\ifx\csname natexlab\endcsname\relax\def\natexlab#1{#1}\fi
\expandafter\ifx\csname bibnamefont\endcsname\relax
  \def\bibnamefont#1{#1}\fi
\expandafter\ifx\csname bibfnamefont\endcsname\relax
  \def\bibfnamefont#1{#1}\fi
\expandafter\ifx\csname citenamefont\endcsname\relax
  \def\citenamefont#1{#1}\fi
\expandafter\ifx\csname url\endcsname\relax
  \def\url#1{\texttt{#1}}\fi
\expandafter\ifx\csname urlprefix\endcsname\relax\def\urlprefix{URL }\fi
\providecommand{\bibinfo}[2]{#2}
\providecommand{\eprint}[2][]{\url{#2}}

\bibitem[{\citenamefont{R.Weht and W.E.Pickett}(1998)}]{weht1}
\bibinfo{author}{\bibnamefont{R.Weht}} \bibnamefont{and}
  \bibinfo{author}{\bibnamefont{W.E.Pickett}}, \bibinfo{journal}{Phys. Rev.
  Lett.} \textbf{\bibinfo{volume}{81}}, \bibinfo{pages}{2502}
  (\bibinfo{year}{1998}).

\bibitem[{\citenamefont{R.Neudert et~al.}(1999)\citenamefont{R.Neudert,
  H.Rosner, S-L.Drehsler, M.Kielwein, M.Sing, Z.Hu, M.Knupfer, M.S.Golden,
  J.Fink, N.Nucker et~al.}}]{neudert1}
\bibinfo{author}{\bibnamefont{R.Neudert}},
  \bibinfo{author}{\bibnamefont{H.Rosner}},
  \bibinfo{author}{\bibnamefont{S-L.Drehsler}},
  \bibinfo{author}{\bibnamefont{M.Kielwein}},
  \bibinfo{author}{\bibnamefont{M.Sing}}, \bibinfo{author}{\bibnamefont{Z.Hu}},
  \bibinfo{author}{\bibnamefont{M.Knupfer}},
  \bibinfo{author}{\bibnamefont{M.S.Golden}},
  \bibinfo{author}{\bibnamefont{J.Fink}},
  \bibinfo{author}{\bibnamefont{N.Nucker}}, \bibnamefont{et~al.},
  \bibinfo{journal}{Phys.Rev.B} \textbf{\bibinfo{volume}{60}},
  \bibinfo{pages}{13413} (\bibinfo{year}{1999}).

\bibitem[{\citenamefont{Y.Mizuno et~al.}(1998)\citenamefont{Y.Mizuno,
  T.Tohyama, S.Maekawa, T.Osafune, N.Motoyama, H.Eisaki, and
  S.Uchida}}]{mizuno1}
\bibinfo{author}{\bibnamefont{Y.Mizuno}},
  \bibinfo{author}{\bibnamefont{T.Tohyama}},
  \bibinfo{author}{\bibnamefont{S.Maekawa}},
  \bibinfo{author}{\bibnamefont{T.Osafune}},
  \bibinfo{author}{\bibnamefont{N.Motoyama}},
  \bibinfo{author}{\bibnamefont{H.Eisaki}}, \bibnamefont{and}
  \bibinfo{author}{\bibnamefont{S.Uchida}}, \bibinfo{journal}{Phys. Rev. B}
  \textbf{\bibinfo{volume}{57}}, \bibinfo{pages}{5326} (\bibinfo{year}{1998}).

\bibitem[{\citenamefont{B.Delley et~al.}(1999)\citenamefont{B.Delley,
  H.B.Braun, B.Roessli, A.Amato, and U.Staub}}]{delley1}
\bibinfo{author}{\bibnamefont{B.Delley}},
  \bibinfo{author}{\bibnamefont{H.B.Braun}},
  \bibinfo{author}{\bibnamefont{B.Roessli}},
  \bibinfo{author}{\bibnamefont{A.Amato}}, \bibnamefont{and}
  \bibinfo{author}{\bibnamefont{U.Staub}}, \bibinfo{journal}{PSI Scientific
  report} \textbf{\bibinfo{volume}{III}}, \bibinfo{pages}{78}
  (\bibinfo{year}{1999}).

\bibitem[{\citenamefont{de~Graaf et~al.}(2002)\citenamefont{de~Graaf,
  de~P.R.~Moreira, Illas, \'{O}.Iglesias, and Labarta}}]{degraaf1}
\bibinfo{author}{\bibfnamefont{C.}~\bibnamefont{de~Graaf}},
  \bibinfo{author}{\bibfnamefont{I.}~\bibnamefont{de~P.R.~Moreira}},
  \bibinfo{author}{\bibfnamefont{F.}~\bibnamefont{Illas}},
  \bibinfo{author}{\bibnamefont{\'{O}.Iglesias}}, \bibnamefont{and}
  \bibinfo{author}{\bibfnamefont{A.}~\bibnamefont{Labarta}},
  \bibinfo{journal}{Phys. Rev. B} \textbf{\bibinfo{volume}{66}},
  \bibinfo{pages}{014448} (\bibinfo{year}{2002}).

\bibitem[{\citenamefont{J.B.Goodenough}(1955)}]{goodenough1}
\bibinfo{author}{\bibnamefont{J.B.Goodenough}}, \bibinfo{journal}{Phys. Rev.}
  \textbf{\bibinfo{volume}{100}}, \bibinfo{pages}{564} (\bibinfo{year}{1955}).

\bibitem[{\citenamefont{J.Kanamori}(1959)}]{kanamori1}
\bibinfo{author}{\bibnamefont{J.Kanamori}}, \bibinfo{journal}{J. Phys. Chem.
  Solids} \textbf{\bibinfo{volume}{10}}, \bibinfo{pages}{87}
  (\bibinfo{year}{1959}).

\bibitem[{\citenamefont{P.W.Anderson}(1963)}]{anderson1}
\bibinfo{author}{\bibnamefont{P.W.Anderson}}, \bibinfo{journal}{Solid State
  Phys.} \textbf{\bibinfo{volume}{14}}, \bibinfo{pages}{99}
  (\bibinfo{year}{1963}).

\bibitem[{\citenamefont{J.W.Lynn et~al.}(1976)\citenamefont{J.W.Lynn,
  G.Shirane, and M.Blume}}]{lynn1}
\bibinfo{author}{\bibnamefont{J.W.Lynn}},
  \bibinfo{author}{\bibnamefont{G.Shirane}}, \bibnamefont{and}
  \bibinfo{author}{\bibnamefont{M.Blume}}, \bibinfo{journal}{Phys. Rev. Lett.}
  \textbf{\bibinfo{volume}{37}}, \bibinfo{pages}{154} (\bibinfo{year}{1976}).

\bibitem[{\citenamefont{R.Hoppe and H.Riek}(1970)}]{hoppe1}
\bibinfo{author}{\bibnamefont{R.Hoppe}} \bibnamefont{and}
  \bibinfo{author}{\bibnamefont{H.Riek}}, \bibinfo{journal}{Z. Anorg. Allg.
  Chem.} \textbf{\bibinfo{volume}{379}}, \bibinfo{pages}{157}
  (\bibinfo{year}{1970}).

\bibitem[{\citenamefont{Ohta et~al.}(1993)\citenamefont{Ohta, Yamauchi, Nanba,
  Motokawa, Kawamata, and Okuda}}]{ohta1}
\bibinfo{author}{\bibfnamefont{H.}~\bibnamefont{Ohta}},
  \bibinfo{author}{\bibfnamefont{N.}~\bibnamefont{Yamauchi}},
  \bibinfo{author}{\bibfnamefont{T.}~\bibnamefont{Nanba}},
  \bibinfo{author}{\bibfnamefont{M.}~\bibnamefont{Motokawa}},
  \bibinfo{author}{\bibfnamefont{S.}~\bibnamefont{Kawamata}}, \bibnamefont{and}
  \bibinfo{author}{\bibfnamefont{K.}~\bibnamefont{Okuda}}, \bibinfo{journal}{J.
  Phys. Soc. Jpn.} \textbf{\bibinfo{volume}{62}}, \bibinfo{pages}{785}
  (\bibinfo{year}{1993}).

\bibitem[{\citenamefont{K.Shreedhar and P.Ganguly}(1988)}]{shreedhar1}
\bibinfo{author}{\bibnamefont{K.Shreedhar}} \bibnamefont{and}
  \bibinfo{author}{\bibnamefont{P.Ganguly}}, \bibinfo{journal}{Inorg. Chem.}
  \textbf{\bibinfo{volume}{27}}, \bibinfo{pages}{2261} (\bibinfo{year}{1988}).

\bibitem[{\citenamefont{K.Okuda et~al.}(1992)\citenamefont{K.Okuda, S.Noguchi,
  K.Konishi, H.Degchi, and K.Takeda}}]{okuda1}
\bibinfo{author}{\bibnamefont{K.Okuda}},
  \bibinfo{author}{\bibnamefont{S.Noguchi}},
  \bibinfo{author}{\bibnamefont{K.Konishi}},
  \bibinfo{author}{\bibnamefont{H.Degchi}}, \bibnamefont{and}
  \bibinfo{author}{\bibnamefont{K.Takeda}}, \bibinfo{journal}{J. Magn. Magn.
  Mater.} \textbf{\bibinfo{volume}{104}}, \bibinfo{pages}{817}
  (\bibinfo{year}{1992}).

\bibitem[{\citenamefont{R.Hoffmann et~al.}(1989)\citenamefont{R.Hoffmann,
  R.Hoppe, and W.Sch{\"a}fer}}]{hoffmann1}
\bibinfo{author}{\bibnamefont{R.Hoffmann}},
  \bibinfo{author}{\bibnamefont{R.Hoppe}}, \bibnamefont{and}
  \bibinfo{author}{\bibnamefont{W.Sch{\"a}fer}}, \bibinfo{journal}{Z. Anorg.
  Allg. Chem.} \textbf{\bibinfo{volume}{578}}, \bibinfo{pages}{18}
  (\bibinfo{year}{1989}).

\bibitem[{\citenamefont{F.Sapi{\~n}a et~al.}(1990)\citenamefont{F.Sapi{\~n}a,
  J.Rodriguez-Carvajal, M.J.Sanchis, R.Ibanez, A.Beltran, and
  D.Beltran}}]{sapina1}
\bibinfo{author}{\bibnamefont{F.Sapi{\~n}a}},
  \bibinfo{author}{\bibnamefont{J.Rodriguez-Carvajal}},
  \bibinfo{author}{\bibnamefont{M.J.Sanchis}},
  \bibinfo{author}{\bibnamefont{R.Ibanez}},
  \bibinfo{author}{\bibnamefont{A.Beltran}}, \bibnamefont{and}
  \bibinfo{author}{\bibnamefont{D.Beltran}}, \bibinfo{journal}{Solid State
  Commun.} \textbf{\bibinfo{volume}{74}}, \bibinfo{pages}{779}
  (\bibinfo{year}{1990}).

\bibitem[{\citenamefont{U.Staub et~al.}(2000)\citenamefont{U.Staub, B.Roessli,
  and A.Amato}}]{staub1}
\bibinfo{author}{\bibnamefont{U.Staub}},
  \bibinfo{author}{\bibnamefont{B.Roessli}}, \bibnamefont{and}
  \bibinfo{author}{\bibnamefont{A.Amato}}, \bibinfo{journal}{Physica B}
  \textbf{\bibinfo{volume}{289-290}}, \bibinfo{pages}{299}
  (\bibinfo{year}{2000}).

\bibitem[{\citenamefont{R.J.Ortega et~al.}(1998)\citenamefont{R.J.Ortega,
  P.J.Jensen, K.V.Rao, F.Sapi{\~n}a, D.Beltran, Z.Iqbal, J.C.Cooley, and
  J.L.Smith}}]{ortega1}
\bibinfo{author}{\bibnamefont{R.J.Ortega}},
  \bibinfo{author}{\bibnamefont{P.J.Jensen}},
  \bibinfo{author}{\bibnamefont{K.V.Rao}},
  \bibinfo{author}{\bibnamefont{F.Sapi{\~n}a}},
  \bibinfo{author}{\bibnamefont{D.Beltran}},
  \bibinfo{author}{\bibnamefont{Z.Iqbal}},
  \bibinfo{author}{\bibnamefont{J.C.Cooley}}, \bibnamefont{and}
  \bibinfo{author}{\bibnamefont{J.L.Smith}}, \bibinfo{journal}{J. Appl. Phys.}
  \textbf{\bibinfo{volume}{83}}, \bibinfo{pages}{6542} (\bibinfo{year}{1998}).

\bibitem[{\citenamefont{S.Patat et~al.}(1991)\citenamefont{S.Patat, D.P.Blunt,
  A.M.Chippendale, and P.G.Dickens}}]{patat1}
\bibinfo{author}{\bibnamefont{S.Patat}},
  \bibinfo{author}{\bibnamefont{D.P.Blunt}},
  \bibinfo{author}{\bibnamefont{A.M.Chippendale}}, \bibnamefont{and}
  \bibinfo{author}{\bibnamefont{P.G.Dickens}}, \bibinfo{journal}{Solid State
  Ionics} \textbf{\bibinfo{volume}{46}}, \bibinfo{pages}{325}
  (\bibinfo{year}{1991}).

\bibitem[{\citenamefont{C.Wilkinson et~al.}(1988)\citenamefont{C.Wilkinson,
  H.W.Khamis, R.F.D.Stansfield, and G.J.McIntyre}}]{wilkinson1}
\bibinfo{author}{\bibnamefont{C.Wilkinson}},
  \bibinfo{author}{\bibnamefont{H.W.Khamis}},
  \bibinfo{author}{\bibnamefont{R.F.D.Stansfield}}, \bibnamefont{and}
  \bibinfo{author}{\bibnamefont{G.J.McIntyre}}, \bibinfo{journal}{J. Appl.
  Crystallography.} \textbf{\bibinfo{volume}{21}}, \bibinfo{pages}{471}
  (\bibinfo{year}{1988}).

\bibitem[{\citenamefont{P.Coppens et~al.}(1965)\citenamefont{P.Coppens,
  L.Leiserowitz, and D.Rabinovich}}]{coppens1}
\bibinfo{author}{\bibnamefont{P.Coppens}},
  \bibinfo{author}{\bibnamefont{L.Leiserowitz}}, \bibnamefont{and}
  \bibinfo{author}{\bibnamefont{D.Rabinovich}}, \bibinfo{journal}{Acta.
  Crystallogr.} \textbf{\bibinfo{volume}{18}}, \bibinfo{pages}{1035}
  (\bibinfo{year}{1965}).

\bibitem[{\citenamefont{P.J.Becker and P.Coppens}(1974)}]{becker1}
\bibinfo{author}{\bibnamefont{P.J.Becker}} \bibnamefont{and}
  \bibinfo{author}{\bibnamefont{P.Coppens}}, \bibinfo{journal}{Acta
  Crystallogr.} \textbf{\bibinfo{volume}{A30}}, \bibinfo{pages}{129}
  (\bibinfo{year}{1974}).

\bibitem[{\citenamefont{P.J.Brown}(1992)}]{itc2}
\bibinfo{author}{\bibnamefont{P.J.Brown}}, \emph{\bibinfo{title}{International
  Tables for Crystallography}} (\bibinfo{publisher}{Kluwer Academic Publishers,
  Dordrecht}, \bibinfo{year}{1992}), vol.~\bibinfo{volume}{C}, chap.
  \bibinfo{chapter}{4.4.5}, p. \bibinfo{pages}{230}.

\bibitem[{\citenamefont{C.Wilkinson}(1968)}]{wilkinson2}
\bibinfo{author}{\bibnamefont{C.Wilkinson}}, \bibinfo{journal}{Phil. Mag.}
  \textbf{\bibinfo{volume}{17}}, \bibinfo{pages}{609} (\bibinfo{year}{1968}).

\bibitem[{\citenamefont{C.Wilkinson}(1973)}]{wilkinson3}
\bibinfo{author}{\bibnamefont{C.Wilkinson}}, \bibinfo{journal}{Acta
  Crystallogr.} \textbf{\bibinfo{volume}{A29}}, \bibinfo{pages}{449}
  (\bibinfo{year}{1973}).

\bibitem[{\citenamefont{C.Wilkinson and E.J.Lisher}(1973)}]{wilkinson4}
\bibinfo{author}{\bibnamefont{C.Wilkinson}} \bibnamefont{and}
  \bibinfo{author}{\bibnamefont{E.J.Lisher}}, \bibinfo{journal}{Acta
  Crystallogr.} \textbf{\bibinfo{volume}{A29}}, \bibinfo{pages}{453}
  (\bibinfo{year}{1973}).

\bibitem[{\citenamefont{O.Halpern and M.H.Johnson}(1939)}]{halpern1}
\bibinfo{author}{\bibnamefont{O.Halpern}} \bibnamefont{and}
  \bibinfo{author}{\bibnamefont{M.H.Johnson}}, \bibinfo{journal}{Phys. Rev.}
  \textbf{\bibinfo{volume}{55}}, \bibinfo{pages}{898} (\bibinfo{year}{1939}).

\bibitem[{\citenamefont{P.J.Brown and J.C.Matthewman}(1993)}]{ccsl}
\bibinfo{author}{\bibnamefont{P.J.Brown}} \bibnamefont{and}
  \bibinfo{author}{\bibnamefont{J.C.Matthewman}}, \bibinfo{type}{Tech. Rep.},
  \bibinfo{institution}{Rutherford Appleton Laboratory} (\bibinfo{year}{1993}).

\bibitem[{\citenamefont{P.J.Brown}(2000)}]{Oformfactor}
\bibinfo{author}{\bibnamefont{P.J.Brown}} (\bibinfo{year}{2000}),
  \bibinfo{note}{private communication}.

\bibitem[{\citenamefont{M.Boehm et~al.}(1998)\citenamefont{M.Boehm, S.Coad,
  B.Roessli, A.Zheludev, M.Zolliker, P.Boni, D.McK.Paul, H.Eisaki, N.Motoyama,
  and S.Uchida}}]{boehm1}
\bibinfo{author}{\bibnamefont{M.Boehm}},
  \bibinfo{author}{\bibnamefont{S.Coad}},
  \bibinfo{author}{\bibnamefont{B.Roessli}},
  \bibinfo{author}{\bibnamefont{A.Zheludev}},
  \bibinfo{author}{\bibnamefont{M.Zolliker}},
  \bibinfo{author}{\bibnamefont{P.Boni}},
  \bibinfo{author}{\bibnamefont{D.McK.Paul}},
  \bibinfo{author}{\bibnamefont{H.Eisaki}},
  \bibinfo{author}{\bibnamefont{N.Motoyama}}, \bibnamefont{and}
  \bibinfo{author}{\bibnamefont{S.Uchida}}, \bibinfo{journal}{Europhys. lett.}
  \textbf{\bibinfo{volume}{43}}, \bibinfo{pages}{77} (\bibinfo{year}{1998}).

\bibitem[{\citenamefont{J.Pierre et~al.}(1998)\citenamefont{J.Pierre, B.Gillon,
  L.Pinsard, and A.Revcolevschi}}]{pierre1}
\bibinfo{author}{\bibnamefont{J.Pierre}},
  \bibinfo{author}{\bibnamefont{B.Gillon}},
  \bibinfo{author}{\bibnamefont{L.Pinsard}}, \bibnamefont{and}
  \bibinfo{author}{\bibnamefont{A.Revcolevschi}}, \bibinfo{journal}{Europhys.
  Lett.} \textbf{\bibinfo{volume}{42}}, \bibinfo{pages}{85}
  (\bibinfo{year}{1998}).

\bibitem[{\citenamefont{M.Matsuda et~al.}(1998)\citenamefont{M.Matsuda,
  K.M.Kojima, Y.L.Uemura, J.L.Zaretsky, K.Nakajima, K.Kakurai, T.Yokoo,
  S.M.Shapira, and G.Shirane}}]{matsuda1}
\bibinfo{author}{\bibnamefont{M.Matsuda}},
  \bibinfo{author}{\bibnamefont{K.M.Kojima}},
  \bibinfo{author}{\bibnamefont{Y.L.Uemura}},
  \bibinfo{author}{\bibnamefont{J.L.Zaretsky}},
  \bibinfo{author}{\bibnamefont{K.Nakajima}},
  \bibinfo{author}{\bibnamefont{K.Kakurai}},
  \bibinfo{author}{\bibnamefont{T.Yokoo}},
  \bibinfo{author}{\bibnamefont{S.M.Shapira}}, \bibnamefont{and}
  \bibinfo{author}{\bibnamefont{G.Shirane}}, \bibinfo{journal}{Phys. Rev. B}
  \textbf{\bibinfo{volume}{57}}, \bibinfo{pages}{11467} (\bibinfo{year}{1998}).

\end{thebibliography}

\end{document}